\documentclass[12pt]{iopart}

\newcommand{\be}{\begin{equation}} 
\newcommand{\ee}{\end{equation}}
\newcommand{\beqn}{\begin{eqnarray}} 
\newcommand{\eeqn}{\end{eqnarray}}

\usepackage{graphicx}
\usepackage{epsfig}
\usepackage{color}
 
\begin{document}
\input epsf.sty

\title{Anomalous coarsening in disordered exclusion processes}

\author{R\'obert Juh\' asz}
\address{Institute for Solid State Physics and Optics, Wigner Research Centre for Physics, H-1525 Budapest, P.O. Box 49, Hungary}
\ead{juhasz.robert@wigner.mta.hu}
\author{G\'eza \'Odor}
\address{Research Centre for Natural Sciences,
Hungarian Academy of Sciences, MTA TTK MFA,
P. O. Box 49, H-1525 Budapest, Hungary}
\ead{odor@mfa.kfki.hu}

\date{\today}

\begin{abstract}
We study coarsening phenomena in three different simple exclusion processes 
with quenched disordered jump rates. 
In the case of the totally asymmetric process, an earlier phenomenological
description is improved, yielding for the time dependence of the length scale  
$\xi(t)\sim t/(\ln t)^2$, which is found to be in agreement with results of
Monte Carlo simulations.   
For the partially asymmetric process, the logarithmically slow coarsening
predicted by a phenomenological theory is confirmed by Monte Carlo
simulations and numerical mean-field calculations. 
Finally, coarsening in a bidirectional, two-lane model with random 
lane-change rates is studied. 
Here, Monte Carlo simulations indicate an unusual
dependence of the dynamical exponent on the density of particles.  
\end{abstract}

\maketitle

\section{Introduction}

Single-file transport is known to be susceptible to
heterogeneities of the track on which the process takes place---one may think
of formation of jams in vehicular traffic \cite{santen} or shocks in the
intracellular traffic of molecular motors on the 
filamentary network \cite{experimental}. 
Such systems are frequently modeled by some variant of the asymmetric simple
exclusion process (ASEP) \cite{mcdonald,spitzer}, where particles jump
stochastically on a lattice and interact through excluding multiple occupations
on lattice sites. 
For this model, in its homogeneous form, many exact results are
available \cite{liggett,schutzreview} and, therefore, it has become a 
paradigmatic model of non-equilibrium systems. 
Motivated by understanding transport in heterogeneous systems,
the ASEP with quenched random jump rates has been the subject 
of ongoing research
\cite{ramaswamy,stanley,tripathy,goldstein,kolwankar,krug2,hs,jsi,jli06,barma,schadschneider,mf}. 
Besides the experimental relevance, the effect of disorder on this model 
is also a challenging problem as general exact solutions are still lacking
here, and the bulk of the results have been obtained by mean-field 
approximations, phenomenological theories based on extreme-value statistics 
and Monte Carlo simulations.  

Since the ASEP can be mapped to a growth process of a $1+1$ dimensional
surface \cite{kpz-asepmap} which is described by
the Kardar-Parisi-Zhang equation \cite{KPZeq}, results for
exclusion processes are also important to understand the fundamentals of
surface physics \cite{barabasi} or directed polymers in random media 
\cite{KPZeq}.
Due to the steady current, 
the correlation length diverges in these systems and, consequently, they 
exhibit a non-equilibrium critical scaling behavior that can be 
classified into universality classes \cite{odor}.
Understanding the effects of disorder on lattice gas models thus 
reveals the behavior of the corresponding interfaces. In particular,
variants of the ASEP with quenched randomness in the hopping rates are 
related to surface growth models with columnar disorder (see \cite{odor}). 

Typically, these models relax toward the steady state 
much slower than the corresponding homogeneous ones; in certain cases, 
the coarsening dynamics are logarithmically slow. 
Owing to this, confirming the predictions of
phenomenological theories by Monte Carlo simulations is a very hard
numerical task. 
As it has been recently shown, inhomogeneous exclusion processes 
can be parallelized and simulated efficiently on a GPU \cite{GPUtexcikk}, 
similar to other interacting systems \cite{preis,weigel,BPP10}. 
In this work, we shall get use of this new technique (among other methods,
such mean-field approximation) in studying the dynamics of three variants of 
the disordered ASEP: 
(i) the disordered totally asymmetric simple exclusion process (DTASEP),
where particles move unidirectionally, 
(ii) the disordered partially asymmetric simple exclusion process (DPASEP), where
particles can move in both directions and 
(iii) a recently introduced bidirectional two-lane model, where the disorder
appears in the lane-changes \cite{J10}. 
We shall see that the above disordered processes with parallel (or synchronous) 
update procedure have essentially the same physical properties as those with
(the most frequently studied) random sequential (or asynchronous) update
procedure. 
In comparing the results with the mean-field description we shall enlighten
the limitations of the latter in case slow modes related to moving domain
walls are present. 

The paper is organized as follows.
In section \ref{tasep} and \ref{pasep}, the coarsening phenomenon in the
DTASEP and in the DPASEP is investigated, respectively.  
Section \ref{twolane} is devoted to the steady state and the coarsening in the
bidirectional two-lane model. 
Finally, results are discussed in section \ref{discussion}. 
Some technical details concerning the parallelization are given 
in the Appendix.

\section{Coarsening in the disordered totally asymmetric simple exclusion process}
\label{tasep}

First, we shall study the disordered totally asymmetric simple exclusion
process defined as follows. A one-dimensional periodic lattice with $L$ sites
is given, on which $N$ particles reside, at most one on each site. 
The TASEP is a continuous-time stochastic process in which particles move 
from site $i$ to site $i+1$ independently 
with rate $p_i$ provided site $i+1$ is empty. 
The jump rates $p_i$ are independent random variables drawn from a bimodal
distribution:   
\begin{equation}\label{bimodal}
f(p_i) = (1-\phi) \delta(p_i-1) + \phi \delta(p_i - r ) \ \ ,
\end{equation}
where $0<r<1$ and $0<\phi<1$. 
Although exact solutions are lacking for this model, the main features of the
steady state and the non-stationary (coarsening) behavior are well understood
by means of a phenomenological description \cite{krug2,barma}. 
This rests on that particles accumulate behind consecutive stretches 
of bonds with rate $p_i=r$ (called bottlenecks) and form a high density domain 
where the local density is greater than $1/2$. 
Provided that the global density $N/L$ is close to $1/2$, 
and the system is initiated in a state with a uniform density, then 
high-density
segments form behind bottlenecks and start to grow as time elapses.
During this process, segments at short bottlenecks lose particles in favor of
those at longer ones until the former gradually vanish. 
Ultimately, provided the system was finite, a single,
macroscopic high density domain accumulates behind the longest 
bottleneck while in the rest of the system the local density is below $1/2$.
So, the system undergoes a coarsening process which is characterized by the
dependence of the typical size of segments $\xi(t)$ on time. 
In a simplified description of this phenomenon, the bottlenecks are treated to
be independent and are characterized by a (length dependent) maximal current.
Furthermore, the essence of the coarsening process is captured if one
concentrates merely on the evolution of length of high density segments     
behind bottlenecks.  
Under these assumptions, the model reduces to a disordered zero-range process
\cite{eh}, where the dynamical exponent $z$ describing the coarsening as 
\be 
\xi(t)\sim t^{1/z}
\ee 
is known to depend on the asymptotics $\rho(w)\sim (w-c)^n$ of the distribution
of jump rates at the lower edge $c$ as $z=(n+2)/(n+1)$ \cite{jain}. 
In the DTASEP, the maximal current through a bottleneck of length $l$ 
is \cite{km} 
\be
J(l)=r/4+O(l^{-1})
\label{Jl}
\ee 
and the distribution of the bottleneck length
is geometric. This yields formally $n=\infty$ and $z=1$. 
In this special limit, non-rigorous arguments based on extreme value
statistics indicate that the
coarsening is not simply linear but logarithmic corrections 
arise as $\xi(t)\sim t/\ln t$ \cite{krug2}. 

Here, we shall reconsider this reasoning and refine it by applying
the statistics of extremes in a more precise way. 
Let us consider a growing high-density segment of length $\xi_i(t)$. Clearly,
the rate of growth of $\xi_i(t)$ is determined the difference $\Delta J_i$ 
between the current $J_{i-1}$ through the bottleneck on the left hand side 
of the growing segment and the current $J_i$ through the bottleneck 
on its right hand side: 
\be 
\dot{\xi_i}(t)\sim\Delta J_i. 
\label{dyn}
\ee
The two bottlenecks are separated by a distance $O(\xi(t))$ and, in this
domain, the bottleneck on the right hand side is the longest one (having
length $l_1$) and the other one is the
second longest one (having length $l_2$). 
Using Eq. (\ref{Jl}), the current difference can be written in terms of
bottleneck lengths as 
\be
\Delta J_i \sim 1/l_2-1/l_1=(l_1-l_2)/l_1l_2.
\label{l12}
\ee
Approximating, for the sake of simplicity, the geometric distribution of bottleneck lengths by the continuous
exponential distribution $P_<(l)=1-e^{-\ln(1/Q)l}$, it is straightforward to
show that  
the most probable values of $l_1$ and $l_2$ among $\xi$ events,   
$l_1^*\approx\ln\xi/\ln(1/Q)$ and 
$l_2^*\approx\ln(\xi/2)/\ln(1/Q)$, respectively, are shifted by a finite
value. Furthermore, the variances of the distributions of $l_1$ and $l_2$ tend
to finite values in the limit $\xi\to\infty$. 
Therefore the limit distribution of  $l_1-l_2$ also has a finite most probable
value and variance and, consequently, the typical current difference 
scales with $\xi$ as 
\be 
\Delta J_i \sim 1/l_1l_2\sim 1/(\ln\xi)^2
\ee
for large $\xi$.
Putting this into Eq. (\ref{dyn}) and integrating yields for the growth of
$\xi(t)$ 
\be 
\xi(t)\sim t/(\ln t)^2
\label{law}
\ee
in leading order for long times.

For measuring the typical length of segments we have made use of a
well-known relationship between the ASEP and a simple one-dimensional surface
growth model where height differences $\Delta h_i=h_{i+1}-h_i$
are related to the occupation number $n_i$ of the ASEP as 
$\Delta h_i = 2n_i-2N/L$ \cite{kpz-asepmap}.  
A usual measure of the width of the surface is given by 
\begin{equation}
W(L,t) = \Bigl[\frac{1}{L} \, \sum_{i=1}^L \,h^2_{i}(t)  -
\Bigl(\frac{1}{L} \, \sum_{i=1}^L \,h_{i}(t) \Bigr)^2
\Bigr]^{1/2} \ .
\label{width}
\end{equation}
In case of the coarsening DTASEP, the equivalent surface consists of
roughly linearly ascending and descending parts corresponding to high and low
density segments, respectively. Therefore the quantity given in
Eq. (\ref{width}) is proportional to the typical value of $\xi(t)$.  

\begin{figure}
\begin{center}
\includegraphics[width=0.4\linewidth]{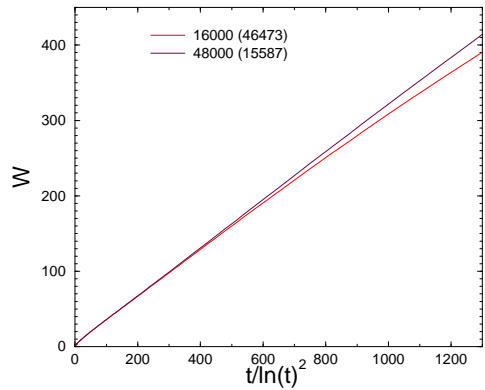}
\caption{\label{W2log}
Dependence of the surface width of the DTASEP on time, obtained by Monte Carlo
simulations with parallel updates for $L=16000, 48000$. 
The number of independent samples are given in parentheses.}
\end{center} 
\end{figure}

\begin{figure}
\begin{center}
\includegraphics[width=0.4\linewidth]{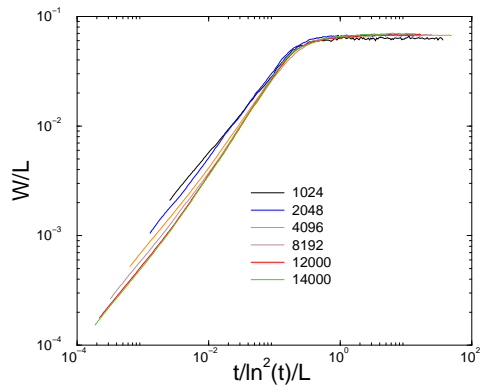}
\caption{\label{col2} Scaling plot of the width $W(L,t)$ obtained for
  different system sizes by Monte Carlo simulations with parallel updates.}
\end{center}
\end{figure}

We performed numerical simulations of the model both with random sequential updates and parallel sub-lattice updates (see
the Appendix) and used the bimodal distribution with parameters 
$\phi=1/2$ and $r=1/4$. The density of particles was $N/L=1/2$. 
We have considered systems with size $L=16000$ and $L=48000$ and measured the
time evolution of the surface width. 
The average of $W(L,t)$ is plotted against $t/(\ln t)^2$ in 
Fig. \ref{W2log}. 
Here, a linear dependence can be seen for not too long times,
where finite size effects are negligible.
Data for finite sizes satisfactorily follow the scaling law 
\be
W(L,t)=L\tilde W(t/(\ln t)^2/L),
\label{logcorr}
\ee
see Fig. \ref{col2}, although, with strong corrections to scaling.

In order to compare the prediction in Eq. (\ref{law}) 
with numerical results more
precisely we have calculated the effective dynamical exponent by 
$1/z_{\rm eff}(t)=d\ln W(t,L)/d\ln t$. 
Due to the logarithmic factor in Eq. (\ref{law}), 
the leading order correction
term in the finite-time effective exponent is slowly decaying, having the form 
\be 
1/z_{\rm eff}(t)=1-2/\ln t + \cdots 
\label{zeff}
\ee
in an infinite system. 
\begin{figure}
\begin{center}
\includegraphics[width=0.6\linewidth]{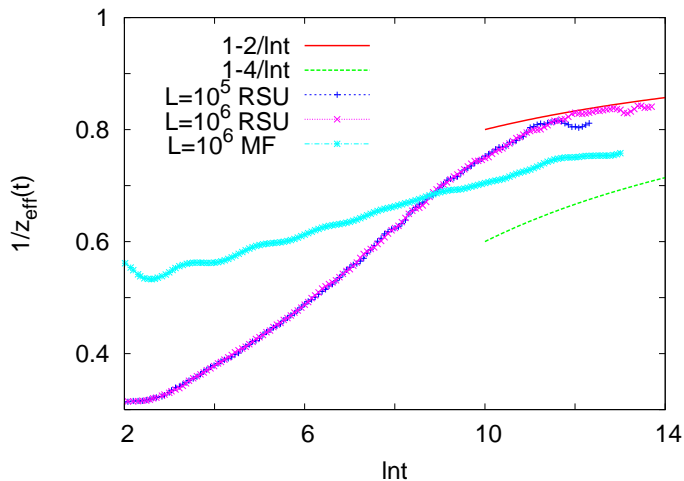}
\caption{\label{wttp} 
Effective dynamical exponents plotted against time obtained by Monte Carlo
simulations with random sequential updates (+,$\times$) and by numerically
solving the numerical mean field equations ($\ast$). The curves correspond to
the predictions of the phenomenological theory.}
\end{center}
\end{figure}
As can be seen in Fig. \ref{wttp}, for moderate time scales  
the effective exponents still deviate from the predictions of the simple
phenomenological theory, nevertheless, the agreement with
Eq. (\ref{zeff}) is satisfactory for long times.     
Due to the slow, logarithmic convergence, 
the estimates on the exponents obtained on small scales differ significantly
from the asymptotic values, see the numerical results in Ref. \cite{QS08}.    

Note, that the simplest scaling hypothesis for roughening surfaces, which is
valid for, among others, the above model in the absence of disorder is of the form
\be
W(L,t)=L^{\alpha}\tilde W(t/L^z)
\label{normal}
\ee
rather than that given Eq. (\ref{logcorr}). 
Here, according to the usual terminology, $\alpha$ is the roughness exponent
characterizing the finite-size scaling of the stationary width of the surface,
i.e. $W(L,t\to\infty)\sim L^{\alpha}$, while the surface growth exponent 
$\beta=z/\alpha$ describes the evolution of the surface width in an
infinite system for long times as  $W(L\to\infty,t)\sim t^{\beta}$.
Hence, apart from the logarithmic correction, 
the surface growth model corresponding to the DTASEP is 
characterized by the exponents $\alpha=\beta=z=1$.

We have also studied the model within a mean-field approximation in which the 
expected values $\langle n_i(t)n_{i+1}(t)\rangle$ in the master equation of
the process are replaced by 
$\langle n_i(t)\rangle\langle n_{i+1}(t)\rangle$. 
Then the master equation turns into the following 
evolution equations for the local densities $\rho_i(t)\equiv\langle n_i(t)\rangle$ 
\be
\frac{d\rho_i}{dt}= p_{i-1}\rho_{i-1}(1-\rho_i)
-p_i\rho_i(1-\rho_{i+1}),  \qquad       i=1,2,\dots,L.
\label{evolution}
\ee
This approximation gives the general features of the ASEP correctly, but,  
since the correlations are neglected, it may yield incorrect 
critical exponents. 
From the point of view of the above phenomenological theory an
important difference is that the finite-size corrections of the current in the
mean-field model are in the order of $l^{-2}$ \cite{QS08}, rather than
$l^{-1}$, see Eq. (\ref{Jl}). 
Repeating the above calculations otherwise unchanged,
leads to a different outcome for the form of the logarithmic correction; 
one obtains namely $\xi(t)\sim t/(\ln t)^4$ and, correspondingly, 
$1/z_{\rm eff}(t)=1-4/\ln t + \cdots$. 
We have solved Eqs. (\ref{evolution}) numerically by using the 4th order
Runge-Kutta method for $10^2$ samples with $L=10^6$.   
The calculated effective exponents are compared with the predictions of the
theory in Fig. \ref{wttp}. As can be seen, on the time-scales of the numerical
investigation a deviation from the predictions is present, which is, however,
decreasing with increasing time. 

\section{Coarsening in the disordered partially asymmetric exclusion process}
\label{pasep}
The dynamics are much different in the quenched disordered partially 
asymmetric exclusion process (DPASEP), where particles can move in both
directions with random rates $p_i$ and $q_i$ to the right and to the left,
respectively, from site $i$ \cite{tripathy,stanley,jsi,barma}. 
Here, due to the large fluctuations of the random potential landscape
\cite{bouchaud}, which is itself a random walk, the displacement of a
single particle increases ultra-slowly as 
\be 
\overline{\langle x^2(t)\rangle}\sim (\ln t)^{2/\psi},
\label{log}
\ee 
with the barrier exponent $\psi=1/2$, 
when the average force $F=\overline{\ln{(p_i/q_{i+1}})}$ 
acting on the particle is zero \cite{sinai}.
Here, $\langle\cdot\rangle$ denotes an average over different stochastic 
histories, whereas the overbar denotes an average over the random transition
rates.
In the following, we shall restrict ourselves to the unbiased case, 
i.e. $F=0$. 
In the presence of many particles (i.e. in the DPASEP), numerical 
simulations showed that
the motion of a tagged particle follows the same law given in Eq. (\ref{log})
\cite{stanley} and, in accordance with the activated dynamics, the
stationary current in finite rings of size $L$ behaves as \cite{jsi} 
\be
-\ln|J(L)|\sim L^{\psi}.
\label{JL}   
\ee 
When the system starts from a homogeneous state it shows a 
coarsening phenomenon
similar to the DTASEP. Queues form at potential barriers, where the local
density is close to one, leaving the rest of the lattice almost completely
empty. As time elapses, queues at smaller barriers gradually dissolve and
particles accumulate behind larger and larger barriers. 
Scaling considerations based on the finite-size scaling behavior of the
current in Eq. (\ref{JL}) lead to that the typical size of high-density
segments increases with time in an infinite system as \cite{jsi}
\be
\xi(t)\sim \ln^{1/\psi}[ t/\ln(t) ].
\label{xit}
\ee
This means that the corresponding surface is characterized formally 
by the exponents $\alpha=1$, $\beta=0$ and $z=\infty$. 
We mention that, in the
presence of a bias ($F>0$), the scaling is normal, i.e. has the form given in
Eq. \ref{normal}, with $\alpha=1$ and with $\beta$ and $z$ continuously
varying with $F$ \cite{jsi}. 

\begin{figure}
\begin{center}
\includegraphics[width=0.4\linewidth]{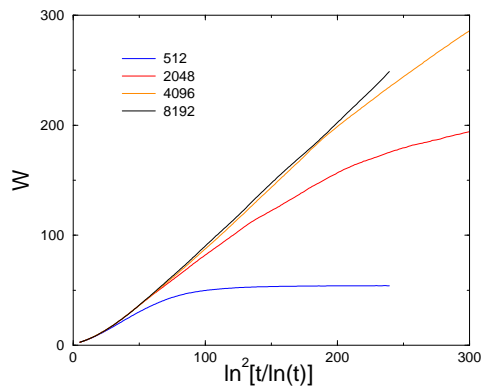}
\caption{\label{wpasep}
Dependence of the surface width of the DPASEP on time, obtained by Monte Carlo
simulations for sizes $L=512, 2048, 4096, 8192$.}
\end{center}
\end{figure}

\begin{figure}
\begin{center}
\includegraphics[width=0.5\linewidth]{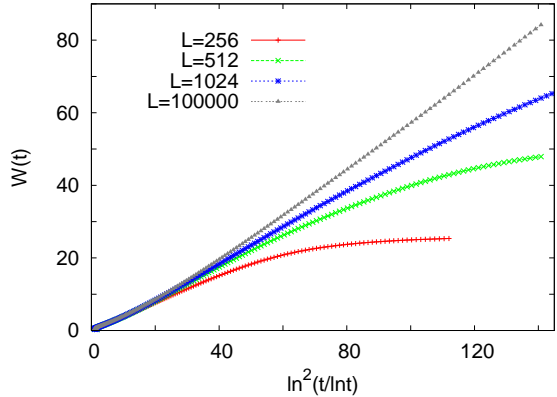}
\caption{\label{wpasepmf}
The same plot as in Fig. \ref{wpasep}, but the data are obtained by solving the dynamical mean-field equations. 
The number of samples was $10^2$.}
\end{center}
\end{figure}

Our aim here is to test the coarsening law in Eq. (\ref{xit}) 
by measuring the dependence of the
corresponding surface width $W(t)$ in Monte Carlo simulations. 
Numerical results obtained with the parallel updating procedure for different
system sizes are shown in Fig. \ref{wpasep}. 
As can be seen, not considering long times where the data are affected by the
finite size of the system, the results are compatible with the law given in
Eq. (\ref{xit}).
We have also considered the model within the mean-field approximation, which
has recently been demonstrated to describe both the stationary behavior and the
dynamics of the model correctly \cite{mf}.
The dynamical mean-field equations read for this model as 
\be
\frac{d\rho_i}{dt}= (p_{i-1}\rho_{i-1}+q_{i}\rho_{i+1})(1-\rho_i)
-[p_i(1-\rho_{i+1})+q_{i-1}(1-\rho_{i-1})]\rho_i.
\label{evolution2}
\ee
In the earlier work, the coarsening length scale has been studied within the
mean-field approximation 
indirectly by measuring the typical distance between adjacent peaks in the
density profile \cite{mf}. 
Here, we have solved the equations Eqs. (\ref{evolution2}) numerically 
and calculated directly the surface width $W(t)$ as a function of time. 
Results shown in Fig. \ref{wpasepmf} are in good agreement with
Eq. (\ref{xit}). 

\section{Dynamics of a bidirectional, two-lane exclusion process with random  lane-change rates}
\label{twolane}

\subsection{Definition of the model and preliminaries}

\begin{figure}[h]
\begin{center}
\includegraphics[width=0.4\linewidth]{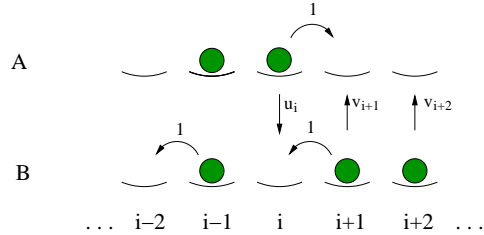}
\caption{
\label{biasep}
Sketch of the bidirectional, two-lane model under study. Allowed transitions
with the corresponding rates are indicated.}
\end{center}
\end{figure}

In the final part of this work, we shall study a recently introduced 
bidirectional, two-lane exclusion process, where the lane-change rates are
random variables \cite{bi,J10}. 
The precise definition of the model is given as follows. 
Two parallel, one-dimensional, periodic lattices are given, each of them with
$L$ sites. Particles move within a lane unidirectionally, following the
dynamical rules of the TASEP with unit jump rates but the directions of motion 
in the two lanes are opposite. In addition two this, particles change lanes
from site $i$ of lane A(B) to the neighboring site of lane B(A) 
with rate $u_i$($v_i$) provided the target site is empty. 
The model and its possible transitions are illustrated in Fig. \ref{biasep}.  
Notice that the direction of motion of particles can be regulated in this model by
tuning the asymmetry in the lane-change rates.
This arrangement was intended to model traffic of molecular motors 
moving along oppositely oriented filaments, which 
can be realized in experiments and may be 
relevant with respect to {\it in vivo} processes \cite{G04}.
A single motor in this environment with homogeneous and
symmetric lane-change rates performs normal diffusion, however, with an
enhanced diffusion coefficient compared to that of the 
symmetric random walk \cite{KL05}. 
Recently, the model has been reinterpreted as a model of motors able to move
bidirectionally along the filament \cite{ashwin}.

We shall consider the disordered version of the above model where the pairs of
rates $(u_i,v_i)$ are i.i.d. random variables. 
In case there is only a single particle in the system, one can introduce an
effective potential, say, in lane A,  
\be 
\Delta U_i\equiv U^A_i-U^A_{i-1}=\ln \frac{1+u_i}{1+v_i},
\label{U}
\ee 
hence, the problem is essentially reduces to a one-dimensional 
random walk in the presence of the above potential field \cite{J10}. 
If $\overline{\Delta U_i}=0$, the time dependence of the displacement follows
Sinai's law given in Eq. (\ref{log}) with $\psi=1/2$. 
What makes the model with many particles difficult is that no
effective potential exists in that case and we have a genuine 
non-equilibrium process. 
An interesting consequence is the following. Having one particle 
in a finite, periodic system where the effective potential is single-valued, 
i.e. $U_1=U_{L+1}$, the expected velocity of the particle will be zero. 
But when putting more particles in the same system, the reduction to an
effective potential field does not work anymore, and the expected velocity
(or current) of particles will be non-zero (except when the sequence of random
rates has a left-right symmetry). 
This {\it interaction induced bias} makes difficult to find a general
criterion for the unbiased point $\overline{J}=0$ in terms of 
the distribution of lane-change rates.
Nevertheless, a sufficient condition for this is that the joint distribution
of rates $P(u,v)$ is symmetric under the interchange of $u$ and $v$. 
In the following we will focus on this case, 
which has not been studied so far. 

The model has exclusively been investigated in the driven phase, where 
$\overline{J}\neq 0$ in finite periodic systems. 
It has been found that the steady state at a bottleneck with
$u>v$ (while in the rest of the system $v>u$), is much different from that of
the one-lane partially asymmetric exclusion process PASEP. 
In the latter, the front of the high-density cluster is localized, 
leading to an exponentially decreasing current with the size $l$ 
of the bottleneck, 
whereas in the former the front performs a symmetric random walk, 
resulting in a larger 
current in the order of $l^{-1}$ \cite{J10}.   
Based on this, the current in the driven phase of the disordered model was
predicted to decay with the system size $L$ as $J(L)\sim 1/\ln L$, 
rather than the much faster algebraic decay in the 
corresponding phase of the one-lane DPASEP.    
Therefore, owing to the collective diffusive motion of 
the high-density cluster, particles overcome barriers more efficiently in the
two-lane model than in the one-lane PASEP.   
We aim here at investigating whether this effect is present at the unbiased
point. Since the phenomenological theory based on independent barriers breaks
down here, we will resort to numerical methods.     

\subsection{Steady state}

A special property of this model, which is due to a symmetry, is that 
if $N=L$, the stationary current is exactly zero for any set of the
lane-change rates, and the local densities fulfill
$\langle n_i^A\rangle=1-\langle n_i^B\rangle$ \cite{J10}. 
If the global density is smaller 
than $1/2$, the stationary current is non-zero 
(but vanishing in the limit $L\to\infty$) and the steady state is segregated,
consisting of a half-filled
domain where approximately $\langle n_i^A\rangle+\langle n_i^B\rangle\approx 1$
and a zero-density domain, where the density is almost zero.
An appropriate measure of typical size of the half-filled domain 
is the corresponding surface width,
but now the surface increment is related to the occupation numbers as 
$\Delta h_i = 2(n_i^A+n_i^B)-2N/L$.
As it is clear from the above picture, 
the surface width in the steady state is proportional to the system size, 
i.e. $W(L,t\to\infty)\sim L$ for $N\neq L$. 
The steady state is, however, much different at half-filling ($N=L$).
In that case, the half-filled phase covers the entire system, thus the 
mean profile $\langle n_i^A\rangle+\langle n_i^B\rangle$ is flat, 
like in a homogeneous system where $u_i=v_i=1$. 
As a comparison, we have investigated the homogeneous system by Monte Carlo
simulations (data not shown) and found the surface width to obey 
the scaling relation  $W(L,t)=L^{1/2}\tilde W(t/L^{2})$ for any density, 
similar to the symmetric simple exclusion process 
belonging to the Edwards-Wilkinson universality class \cite{EW}.
Thereby one would expect the surface width in the disordered model for 
$N=L$ to be reduced, compared to that at other densities. 
Surprisingly, numerical simulations still indicate $W(L,t\to\infty)\sim L$, 
which refers to large fluctuations.
The reason for this is the following.
Besides the half-filled state, there are two other states, 
in which the current is trivially zero: the empty and the full lattice. 
In a half-filled system ($N=L$), let us imagine a state consisting 
of half-filled segments with $\langle n_i^A\rangle=1-\langle n_i^B\rangle$, as
well as empty and fully occupied segments. 
In the homogeneous model, this state will not be stable (i.e. a steady state) 
since particles penetrate
into the empty segments, just as holes into fully occupied segments and move
there with a finite velocity for $u\neq v$, 
leading to that these segments ultimately
dissolve and the whole system is occupied by the half-filled phase. 
In a disordered system, however, the half-filled phase is unstable.  
Here, fully occupied and empty segments are able to
form spontaneously due to inhomogeneities and live for long times.     

\begin{figure}[h]
\begin{center}
\includegraphics[width=0.7\linewidth]{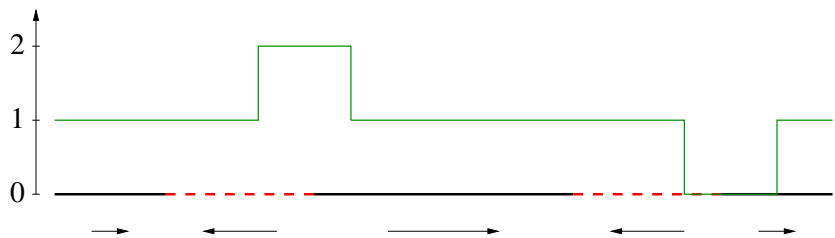}
\caption{
\label{twobarrier}
Snapshot of the density profile $\langle n_i^A\rangle+\langle
n_i^B\rangle$ (green line) 
in a system consisting of two reverse bias regions (denoted by the
dashed red sections). The arrows indicate the direction of motion of
particles in the empty lattice and holes in the occupied lattice.}
\end{center}
\end{figure}

To illustrate this phenomenon, let us consider a caricature of a disordered
system, which contains just two extended regions (called reverse bias regions)
of length $l$, where $u>v$, and which are far from each other, 
while in the rest of the system $v>u$, see Fig \ref{twobarrier}.
An interesting property of the model under study 
is that a particle in an otherwise empty 
region and a hole in an otherwise occupied region move in the same direction;
from left to right (from right to left) if $v>u$ ($u>v$).
Let us assume, that a small empty segment appears around the right end point
of one of the reverse bias regions. 
This segment (or condensate of holes) is stable, as particles, 
penetrating into this segment from the surrounding half-filled
phase (either from left or from right) move preferably away from the end
point, i.e. they are forced back into the half-filled phase. So the domain
walls, separating the empty segment from the surroundings are stable.    
Of course, due to the conservation of particles, a fully
occupied segment must emerge at the right end point of the other reverse bias
region, which will be stable, as well, from similar reasons given above.  
Although the mean particle current through the half-filled phase in between
the two reverse bias regions is zero, its fluctuations result in 
that the size of empty and fully occupied segments fluctuates (in a correlated way due to
particle conservation), as well. The empty (or the fully occupied) region can
extend over a length of $l$ on both sides of the right end point 
of the reverse bias region, independently. 
These correlated fluctuations of the size of particle (and hole) condensates
at the two reverse bias regions amount to a net displacement of particles from
one reverse bias region to the other one and lead to fluctuations of the
corresponding surface. 

Concerning the inhomogeneous model, it can be divided into effective forward and reverse bias
regions, where for the majority of links $v_i>u_i$ and  $v_i<u_i$ hold,
respectively. 
Therefore, similar fluctuations occur as in the above simplified system.
The largest effective reverse bias region is macroscopic in the unbiased case 
\cite{J10}, hence the surface width must scale linearly with 
the system size, in accordance with the observations. 

\subsection{Coarsening}

When the system is started from a homogeneous initial state, 
a coarsening phenomenon can be observed, which is analogous to that in the
previous two models if the system is not at half-filling. 
In this case, the system consists of 
half-filled and empty (fully occupied) segments if $N<L$ ($N>L$), 
the typical size of which is growing in time.
This picture is qualitatively different for $N=L$, nevertheless, starting the
system from a state with $W(t=0)=0$, the surface width must increase in time
also at half-filling.  

\begin{figure}[h]
\begin{center}
\includegraphics[width=0.6\linewidth]{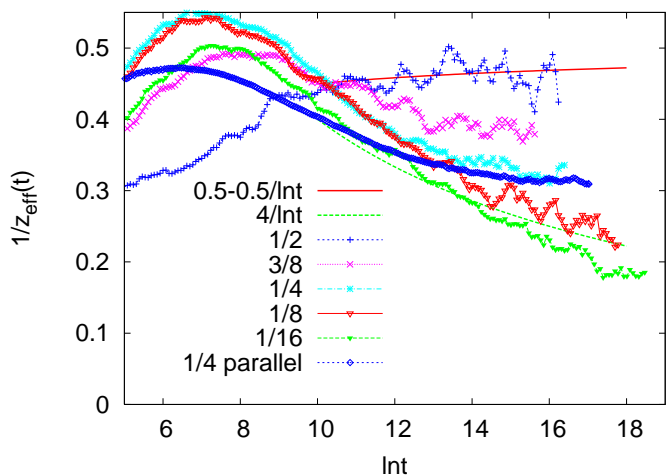}
\caption{
\label{wtltp}
Effective dynamical exponents of coarsening in the bidirectional, two-lane
model measured in Monte Carlo simulations at different global
densities $N/2L=1/2,3/8,1/4,1/8,1/16$. 
Simulations have been perfomed with the random sequential update, 
except one of the data sets, which
was obtained by the parallel update for $L=48000$ and $10^2$ samples.}
\end{center}
\end{figure}

We have performed Monte Carlo simulations for systems 
of size $L=2\cdot 10^5$, starting from a regular configuration
where $W(t=0)=0$ and measured the evolution of the surface width for not too
long times, so that finite-size effects are negligible. 
We have considered different global densities $N/2L=1/2,3/8,1/4,1/8,1/16$ and
several samples for each density.   
The effective dynamical exponents calculated from the average surface width
are plotted against time in Fig. \ref{wtltp}. 
As can be seen in the figure, at half-filling, the effective dynamical
exponent seems to tend to $0.5$ but very slowly, possibly with a logarithmic
correction. The solid curve fitting well to the data
corresponds to the form $W(t)\sim (t/\ln t)^{1/2}$. 
Below half-filling, the effective exponent $1/z_{\rm eff}(t)$ is decreasing
with $t$ for long times. 
For large enough densities, such as $3/8$ and $1/4$, it seems to
tend to finite limiting values which decrease with the density. 
For smaller densities the effective exponents $1/z_{\rm eff}(t)$
decrease more quickly and from 
the data obtained at finite time scales it is difficult to decide whether
they tend to finite values or to zero. The latter case may correspond to a
logarithmic coarsening law $W(t)\sim [\ln (t/t_0)]^{1/\psi}$ with some 
finite barrier exponent $\psi$. 
As a comparison, the dependence of the effective exponent $1/z_{\rm eff}(t)$
on time in this case with $\psi=0.25$ and $t_0=1$ is plotted in the figure.   

\subsection{The mean-field model}

\begin{figure}[h]
\begin{center}
\includegraphics[width=0.45\linewidth]{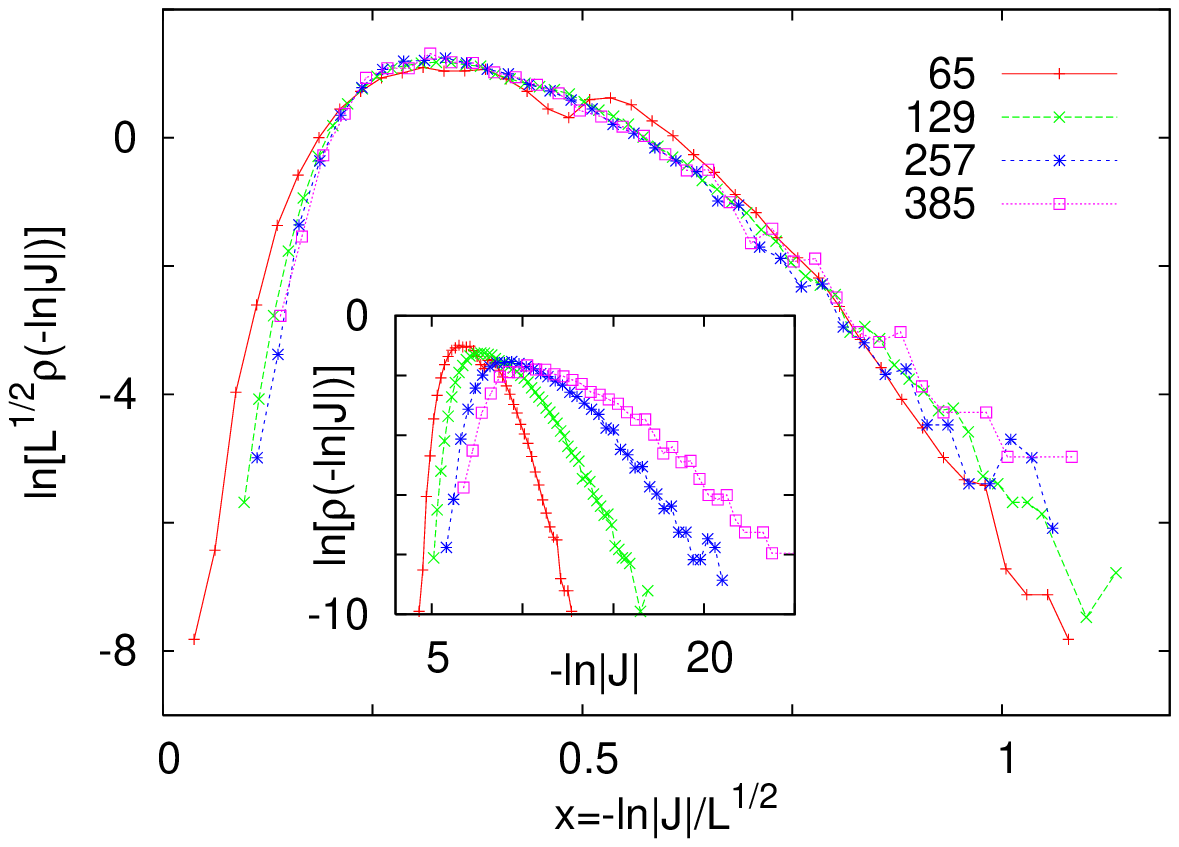}
\includegraphics[width=0.45\linewidth]{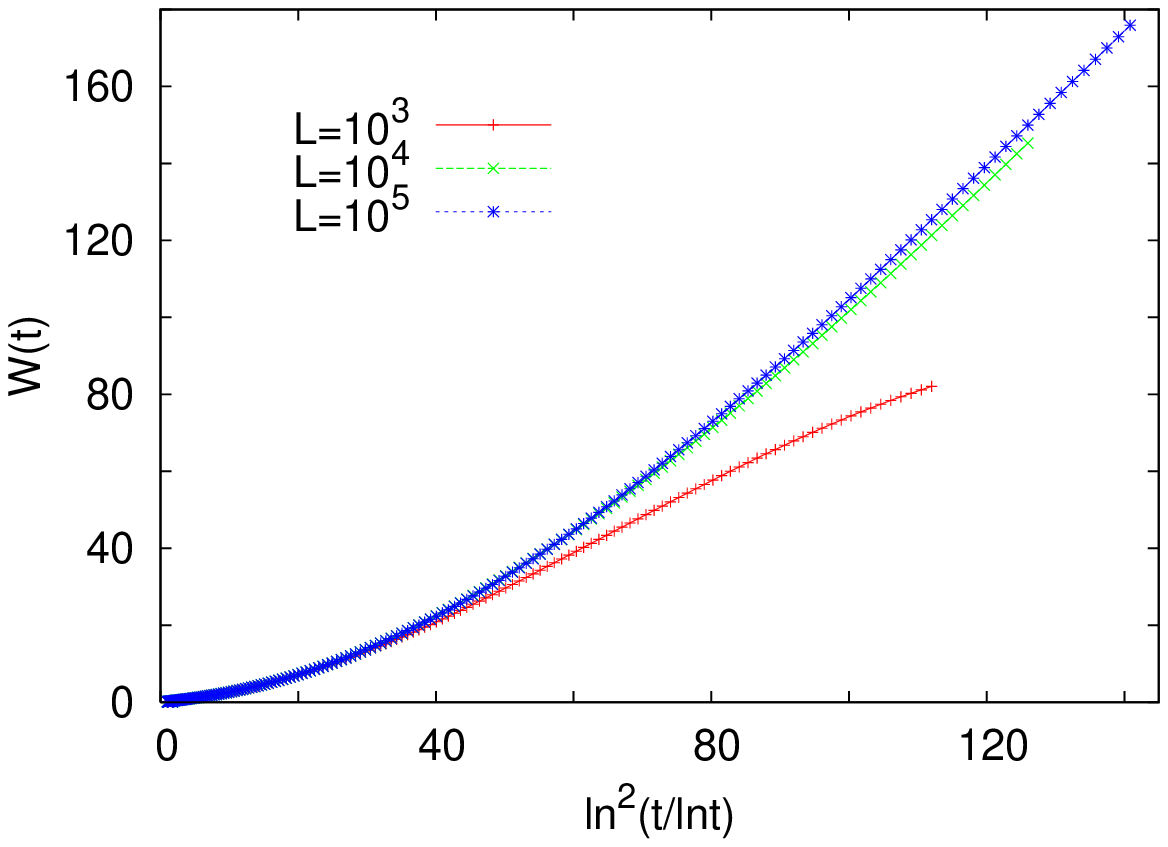}
\caption{
\label{dtlmf}
Left: Scaling plot of the distribution of the logarithm of 
the stationary current obtained for
different sizes by the mean-field approximation. 
The unscaled data are shown in the inset.
Right: Time-dependence of the surface width in the mean-field two-lane model. 
}
\end{center}
\end{figure}

We have also studied the above model within the mean-field approximation. 
Here, one has the following evolution equations for the local densities $\rho_i(t)$ and $\pi_i(t)$ in lane A and B, respectively: 
\beqn
\frac{d\rho_i}{dt}= \rho_{i-1}(1-\rho_i)-\rho_i(1-\rho_{i+1})
+v_i\pi_i(1-\rho_i)-u_i\rho_i(1-\pi_i)
 \nonumber \\
\frac{d\pi_i}{dt}= \pi_{i+1}(1-\pi_i)-\pi_i(1-\pi_{i-1})
-v_i\pi_i(1-\rho_i)+u_i\rho_i(1-\pi_i),
\label{tlmfdyn}
\eeqn
for $i=1,2,\dots,L$.
The steady state of the mean-field model is, however, much different from that 
of the original one, as can be seen from the steady state of the homogeneous
model containing a reverse bias region. The mean-field treatment cannot
describe the diffusive motion of the half-filled particle cluster
mentioned in the previous section. Instead, the front of this cluster 
is pinned in the middle of the reverse bias region. 
This leads to an exponentially decreasing current
with the length of the reverse bias region, and to Eq. (\ref{JL}) for the
finite-size scaling of the current and to Eq. (\ref{xit}) for the coarsening
dynamics in the disordered model. 
We have calculated the stationary current at a density $N/2L=1/4$ 
for different system sizes by
solving Eqs. (\ref{tlmfdyn}) in $10^4$ samples for each $L$.
The distribution of the current shown in Fig. \ref{dtlmf} indeed follows 
the law given in Eq. (\ref{JL}). 
In addition to this, we have followed the evolution of the surface width in
large systems at the same density. The time dependence of the surface width
averaged over a few samples is again in agreement with the prediction in
Eq. (\ref{xit}), as can be seen in Fig. \ref{dtlmf}.       

\section{Discussion}
\label{discussion}

In this work, we have considered different exclusion processes, in which the
disorder in the transition rates results in anomalous dynamics compared to the
clean model. 
For the DTASEP, improving an earlier phenomenological model \cite{krug2}, 
we have obtained
the form of logarithmic correction in the coarsening law, which is found to be
compatible with results of Monte Carlo simulations. 
In the DPASEP, we have confirmed the earlier predicted logarithmically slow
coarsening dynamics by numerical simulations. 

Then we have studied a bidirectional, two-lane exclusion process with
disordered lane-change rates, the dynamics of which are less known than 
in the case of the previous two models. 
In the absence of a potential, the steady state and the non-stationary 
behavior are
non-trivial and are much different compared to those of the disordered 
partially asymmetric one-lane model. 
In the steady state, the disorder induces formation of particle 
and hole condensates of
fluctuating size, which leads to anomalously large fluctuations of the
equivalent surface.
The fact that particle clusters accumulating at bottlenecks 
are delocalized makes possible a faster coarsening than in the DPASEP. 
Numerical simulations show that for particle densities not far from $1/2$, 
the coarsening is described by
power-laws (apart from possible logarithmic corrections) 
with density-dependent dynamical exponents.
For the particular case of the half-filled lattice, 
the dynamics are found to be characterized roughly by the
exponents $\alpha=1$, $\beta=0.5$ and $z=2$, again with possible logarithmic
corrections.
For smaller densities, 
the inverse of the effective dynamical exponent seems to decrease
constantly, but it is not possible to decide from the numerical data whether
it tends to finite values.  

The slowing of the dynamics with decreasing density 
is consistent with that the motion of a single particle is ultra-slow,
characterized formally by an infinite dynamical exponent. 
Note, however, that approaching to the zero density through positive
densities, i.e. performing $N/2L\to 0$ in an infinite system may give  
a limit dynamical law that is different from that of the single particle.   

It is worth mentioning that
even for densities close to $1/2$ the asymptotic region with a presumably
finite dynamical exponent sets in only after a very long transient. 
In the transient, the coarsening, as well as the finite-size scaling of the
stationary current (not shown) can be well described 
by an actived scaling with some effective barrier exponent $\psi$, 
that is smaller than $1/2$.   
Probing the dynamics of the model 
by finite-size scaling of the distribution of the stationary current is hard since, 
due to the long transient, large system sizes are needed and  
samples in the small-current tail of the distribution have extemely long
relaxation times.     
The mean-field version of the above model, where particle
clusters at bottlenecks are localized, shows 
logarithmically slow coarsening formally with an infinitely large dynamical
exponent. 
So we can conclude, that the slow modes related to the delocalized particle
clusters near barriers are important with respect to the dynamics of
coarsening. 

In Table \ref{tabla}, we have summarized the finite-size scaling behavior of 
the stationary current in  disordered variants of the ASEP investigated here or earlier.
\begin{table}
\caption{Comparison of the finite-size scaling of the stationary current in
  the one-lane DPASEP and the bidirectional, two-lane model with disordered
  lane-change rates. 
\label{tabla}}
\begin{center}
\begin{tabular}{|l|r|r|}
\hline
type of model	& one-lane  		& two-lane \\
\hline
single barrier of length $l$& $J \sim e^{-c l}$		& $J \sim 1/l$ \\
biased ($F>0$)	& $J \sim L^{-a(F)}$ 	 	& $J \sim 1/\ln(L)$ \\
unbiased ($F=0$)&$-\ln|J|\sim L^{1/2}$ 	& density dependent \\
\hline
\end{tabular}
\end{center}
\end{table}	  
Interesting extensions of these investigations would be the 
study of dynamics in multi-lane transport systems or 
in the disordered ASEP in higher dimensions \cite{asepddcikk}.

\section{Acknowledgments}

This paper was supported by the J\'anos Bolyai Research Scholarship of the
Hungarian Academy of Sciences (RJ), by 
the Hungarian National Research Fund OTKA under grant nos. T77629, K75324, 
and by the bilateral German-Hungarian exchange program DAAD-M\"OB under grant
nos. 50450744, P-M\"OB/854. 
The authors thank NVIDIA for supporting the project with high-performance
graphics cards within the framework of Professor Partnership.

\appendix
\section{}
\begin{figure}[h]
\begin{center}
\includegraphics[width=0.5\linewidth]{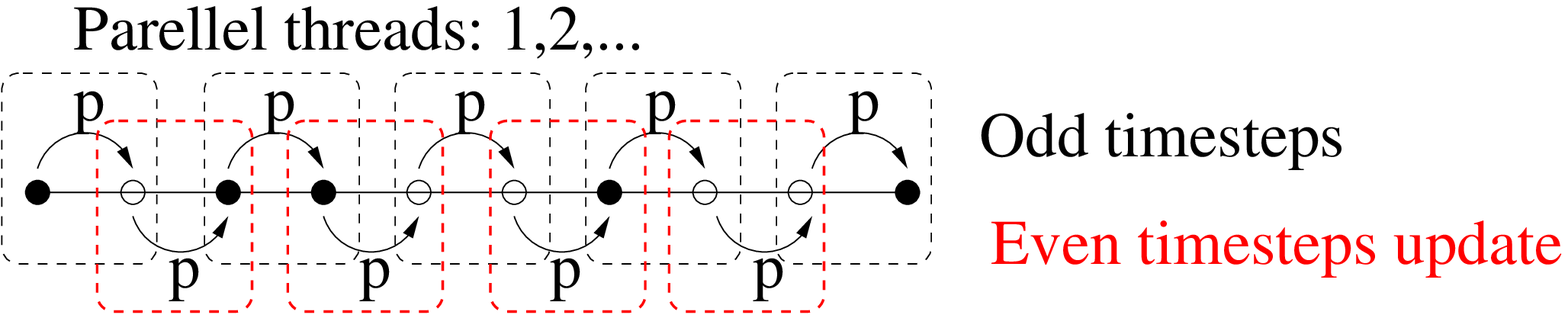}
\caption{Illustration of the parallel, two-sub-lattice update 
applied for the DTASEP}
\label{parasep}
\end{center}
\end{figure}

We have performed the Monte Carlo simulations with parallel updates according
to the following scheme. 
In the case of the DTASEP, we used a parallel two-sub-lattice update shown 
in Fig. \ref{parasep}. Here, the lattice is divided into even and odd 
sub-lattices, which are updated alternately.
In the case of the DPASEP, a similar two-sub-lattice update has been applied
as given above.  

In the case of the bidirectional, two-lane model with parallel update 
each Monte Carlo step consists of the following operations:
(i) even sub-lattice updates of the lanes $A$ and $B$, 
with probabilities $1/2$, 
(ii) lane changes with probabilities $u_i$ and $v_i$,  
(iii) odd sub-lattice updates of the lanes $A$ and $B$
with probabilities $1/2$,
(iv) lane changes with probabilities $u_i$ and $v_i$. 
Further details can be found in Ref. \cite{GPUtexcikk}.



\section*{References}
\begin{thebibliography}{99}

\bibitem{santen} Chowdhury D, Santen L, and Schadschneider A 
2000 {\it Phys. Rep.} {\bf 329} 199

\bibitem{experimental}
Konzack S, Rischitor P E, Enke C and Fischer R 
2005 {\it Mol. Biol. Cell} {\bf 16} 497;
Nishinari K, Okada Y, Schadschneider A and Chowdhury D 
2005 {\it Phys. Rev. Lett.} {\bf 95} 118101;
Greulich P, Garai A, Nishinari K, Schadschneider A and Chowdhury D 
2007 {\it Phys. Rev. E} {\bf 75} 041905


\bibitem{mcdonald} MacDonald C T, Gibbs J H and Pipkin A C 1968 
{\it Biopolymers} {\bf 6} 1

\bibitem{spitzer} 
Spitzer F 1970 {\it Adv. Math.} {\bf 5} 246 


\bibitem{liggett} Liggett T M 1999 {\it Stochastic interacting systems:
  contact, voter, and exclusion processes} (Berlin, Springer)

\bibitem{schutzreview} Sch\"utz G M 2001 in {\it Phase  Transitions and
    Critical Phenomena}, vol. 19, edited by Domb C and Lebowitz J L (Academic, San Diego)

\bibitem{ramaswamy} 
Ramaswamy R, Barma M 1987 {\it J. Phys. A: Math. Gen.} {\bf 20} 2973

\bibitem{stanley}
Koscielny-Bunde E, Bunde A, Havlin S, and Stanley H E
1988 {\it Phys. Rev. A} {\bf 37} 1821

\bibitem{tripathy} Tripathy G, Barma M 1997 {\it Phys. Rev. Lett.} {\bf 78}
  3039; 1998 {\it Phys. Rev. E} {\bf 58} 1911

\bibitem{goldstein}
Goldstein S, Speer E R 1998 {\it Phys. Rev. E} {\bf 58} 4226

\bibitem{kolwankar}
Kolwankar K M, Punnoose A 2000 {\it Phys. Rev. E} {\bf 61} 2453

\bibitem{krug2}
Krug J, 2000 {\it Braz. J. Phys.} {\bf 30} 97

\bibitem{hs}
Harris R J, Stinchcombe R B 2004 {\it Phys. Rev. E} {\bf 70} 016108

\bibitem{jsi}
Juh\'asz R, Santen L and Igl\'oi F 2005 {\it Phys. Rev. Lett.} {\bf 94} 010601;
2006 {\it Phys. Rev. E} {\bf 74} 061101

\bibitem{jli06} Juh\'asz R, Lin Y-C, Igl\'oi F 2006 {\it Phys. Rev. B} {\bf 73} 224206

\bibitem{barma} 
Barma M 2006 {\it Physica A} {\bf 372} 22

\bibitem{schadschneider}
Greulich P, Schadschneider A 2008 {\it J. Stat. Mech.} P04009

\bibitem{mf}
Juh\'asz R 2011 {\it J. Stat. Mech.} P11010

\bibitem{kpz-asepmap}
Plischke M, R\'acz Z and Liu D 1987 {\it Phys. Rev. B} {\bf 35} 3485

\bibitem{KPZeq}
Kardar M, Parisi G and Zhang Y 1986 {\it Phys. Rev. Lett.} {\bf 56} 889

\bibitem{barabasi}
Barab\'asi A L and Stanley H E 1995 {\it Fractal Concepts in Surface Growth}
Cambridge University Press, Cambridge

\bibitem{odor}
\'Odor G 2008 {\it Universality in Nonequilibrium Lattice Systems} World Scientific; 
2004 {\it Rev. Mod. Phys.} {\bf 76} 663

\bibitem{GPUtexcikk} Schulz Henrik, \'Odor G\'eza, \'Odor Gergely, 
Nagy F. M\'at\'e  2011 {\it Comp. Phys. Comm.} {\bf 182} 1467 

\bibitem{preis} 
Preis T at al 2009  {\it J. of Comp. Phys.} {\bf 228} 4468

\bibitem{weigel} Weigel M 2012 {\it  J. of Comp. Phys.} {\bf 231} 3064.

\bibitem{BPP10} Bernaschi M, Parisi G, Parisi L arXiv:1006.2566v1

\bibitem{J10} 
Juh\'asz R 2010 {\it J. Stat. Mech.} P03010 

\bibitem{eh}
Evans M R and Hanney T 2005 {\it J. Phys. A: Math. Gen.} {\bf 38} R195

\bibitem{jain}
Jain K and Barma M 2003 {\it Phys. Rev. Lett.} {\bf 91} 135701

\bibitem{km}
Krug J and Meakin P 1990 {\it J. Phys. A: Math. Gen.} 23 L987

\bibitem{QS08} Queiroz S L A and Stinchcombe R B 2008 
{\it Phys. Rev. E} {\bf 78} 031106

\bibitem{bouchaud} Bouchaud J P, Georges A 1990 {\it Phys. Rep.} {\bf 195} 127


\bibitem{sinai}
Sinai Ya.G. 1982 {\it Theory Probab. Appl.} {\bf 27} 256


\bibitem{bi}
Juh\'asz R 2007 {\it Phys. Rev. E} {\bf 76} 021117

\bibitem{G04} Gross S P 2004 {\it Phys. Biol.} {\bf 1} R1

\bibitem{KL05} Klumpp S and Lipowsky R 2005 {\it Phys. Rev. Lett.} {\bf 95} 268102

\bibitem{ashwin}
Ashwin P, Lin C, Steinberg G 2010  {\it Phys. Rev. E} {\bf 82} 051907 

\bibitem{EW}
Edwards S F, Wilkinson D R 1982 {\it Proc. R. Soc.} {\bf 381} 17

\bibitem{asepddcikk}
\'Odor G, Liedke B, and Heinig K-H 2010 {\it Phys. Rev. E} {\bf 81} 031112


\end{thebibliography}
\end{document}